# Transfer of large-scale perovskite single-crystalline thin films onto arbitrary substrates

Yu-Hao Deng[1*]

[1] Academy for Advanced Interdisciplinary Studies, Peking University, Beijing, China

* Correspondence should be addressed to yuhaodeng@pku.edu.cn

**Abstract**

Metal halide perovskites single-crystalline thin films (SCTFs) have recently emerged as promising materials for the next generation of optoelectronic devices due to their superior intrinsic properties. However, it is still challenging to transfer and integrate them with other functional materials for hybrid multilayer devices because of their substrate-dependent growth. Herein, a method that allows the SCTFs to be transferred with high quality onto arbitrary substrates has been reported. By introducing hydrophobic treatment to the growth substrates, the adhesion between SCTFs and growth substrates is reduced. Meanwhile, anti-solvent intercalation technique is used to peel off SCTFs from the growth substrates intactly. Finally, centimeter-scale perovskite SCTF has been successfully transferred to target substrate. This work opens up a new route to transfer large-scale perovskite SCTFs, providing a platform to widen the applications of perovskite SCTFs in large-scale hybrid multilayer optoelectronic devices.

**Introduction**

Metal halide perovskites have attracted intense interest in the scientific community due to their superior intrinsic properties, which render them promising materials for the next generation of optoelectronic devices, such as solar cells, photodetectors, light-emitting diodes (LED), lasers and X-ray imaging [1-8]. Compared with polycrystalline materials, grain-boundary-free perovskites single-crystalline thin films (SCTFs) exhibit lower trap-state density, higher carrier mobility, longer diffusion lengths and better stability, which are supposed to maximize the performance of solar cells and other optoelectronic devices [9-12]. However, the major barrier for SCTFs in further practical applications is their substrate-dependent growth [13, 14], which remains challenging to transfer and integrate SCTFs with other functional materials for hybrid multilayer optoelectronic devices. Wang et al. reported a transfer method of perovskite nanocrystals by polydimethylsiloxane (PDMS) stamps, wherein the stronger adhesion between perovskite and target substrate drive nanocrystals drop from PDMS stamps [15]. However, this method is limited by substrate adhesion and cannot transfer large-scale perovskite SCTFs by avoiding crystal fragmentation. Recently, Ding et al. developed a wet transferring approach using polystyrene (PS) as carrier layer [16]. PS was spin-coated onto perovskite and then the PS/perovskite was transferred onto target substrate. Finally, the PS was dissolved using organic solvents, leaving the perovskite SCTF on the substrate. This PS-carrier method improves the applicability of target substrates but still suffers from crystal fragmentation of large-scale perovskite SCTFs due to tension deformation of polymer layer when PS/perovskite is peeled off from original substrates. Therefore, a transferring strategy to intactly transfer large-scale perovskite SCTFs onto other functional materials is highly urgent.

**Results and Discussion**

Perovskite SCTFs possess relatively more fragile property than other solar cell materials and highly substrate-dependent growth, which makes them easily damaged during the transfer process [17]. Thus, reducing the adhesion between growth substrates and films is the guarantee of the successful transfer. Hydrophobic treatment for substrates has been demonstrated to be effective in reducing the adhesion between grown materials and substrates [7, 15, 18]. Particularly, the hydrophobic treatment on substrates can reduce nucleation density and accelerate the growth rate of crystals, providing beneficial conditions for the growth of large-scale SCTFs [6, 7, 19, 20]. Fig. 1 shows the schematic illustration of processes. The surfaces of two glass substrates are treated to be hydrophobic by dimethyldimethoxysilane (DMDMS) and put face-to-face to create a space-confined gap. The perovskite material nucleates and grows continuously, and finally forms the SCTF in the gap [7]. Hydrophobic treatment lowers the surface energy of substrates and reduces the adhesion between SCTFs and growth substrates. After the perovskite SCTF is completely grown, the glass-perovskite-glass structure is immersed in the anti-solvent (dichloromethane). Subsequently, the SCTF falls off from substrates automatically due to the intercalation effect of dichloromethane molecules. Finally, the free-standing SCTF is picked up by filter paper, dried naturally, and then transferred to target substrates. Due to the absence of devastating stress, SCTFs keep intact in the process.

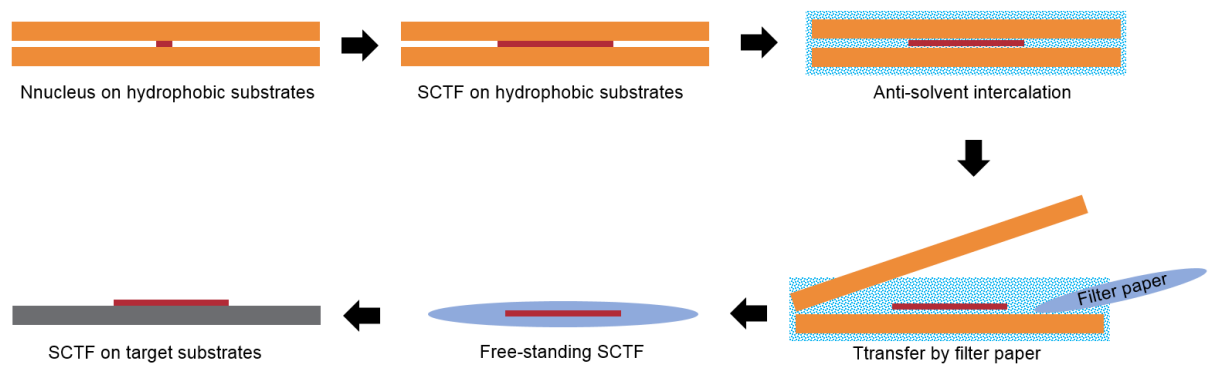

**Fig. 1** Schematic illustration of perovskite SCTF transfer. The SCTF forms in the gap between two hydrophobic glasses by space-confined method. Subsequently, anti-solvent intercalation technique is used to peel off SCTFs from the growth substrates. Finally, the SCTF is picked up by filter paper from the anti-solvent solution, and then transferred to arbitrary substrates.

Fig. 2A shows the MAPbBr$_3$ perovskite SCTF on growth substrate. Although hydrophobic treatment to the substrate has greatly reduced the adhesion between the SCTF and growth substrate, the film still cannot be mechanically removed directly or it will be broken. Fig. 2B shows the SCTF has been transferred onto filter paper after anti-solvent intercalation. In this case, the adhesion of the SCTF to the filter paper is almost gone. Then, the SCTF is free-standing (Fig. 2C) and can be transferred on any other substrates (Fig. 2D).

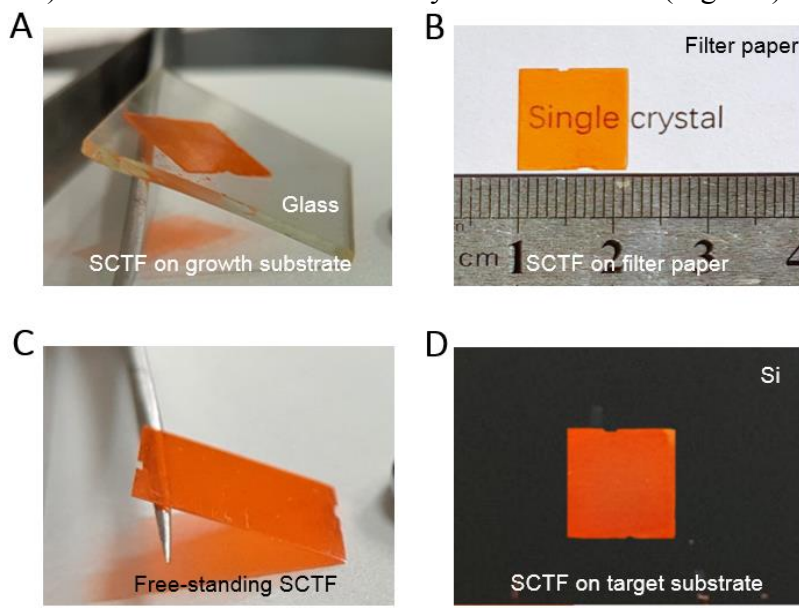

**Fig. 2** The SCTF on different substrates. (A) The MAPbBr3 perovskite SCTF grows on a hydrophobic glass substrate. (B) The SCTF is transferred on filter paper. (C) Free-standing SCTF. (D) The SCTF is transferred onto Si substrate.

Fig. 3A, B show the cross-sectional optical and SEM photograph of the grown MAPbBr$_3$ SC-TF. Step profiler further demonstrates thickness of the SC-TF is 10.7 ± 0.3 μm and the length is 1.15 cm, giving an aspect ratio of 1074 (Fig. 3C). X-ray diffraction (XRD) shows the SC-TF is single crystal with cubic phase and the crystal orientation of the film in vertical direction is (001) (Fig. 3D). The band gap of the SC-TF is calculated as 2.24eV in the absorption spectrum and the steady-state PL spectrum shows the emission peak located at 546 nm with a FWHM of 22 nm (Fig. 3E). As shown in Fig. 3F, the trap density of the SC-TF is measured with the space-charge-limited current analysis under different biases. The trap density is obtained from $n_{trap} = \frac{2\varepsilon_r \varepsilon_0 V_{TFL}}{ed^2} = 1.6 \times 10^{11}$ cm$^{-3}$, where $d$=11 μm is the thickness of the film, $\varepsilon_r = 22.5$ is the relative dielectric constant, and $\varepsilon_0$ is the vacuum permittivity. The trap density of MAPbBr$_3$ SC-TF is 6 orders of magnitude lower than polycrystalline films fabricated by spin-coating methods [21]. The above evidences prove that the transferred SC-TFs have superior optical and electrical properties.

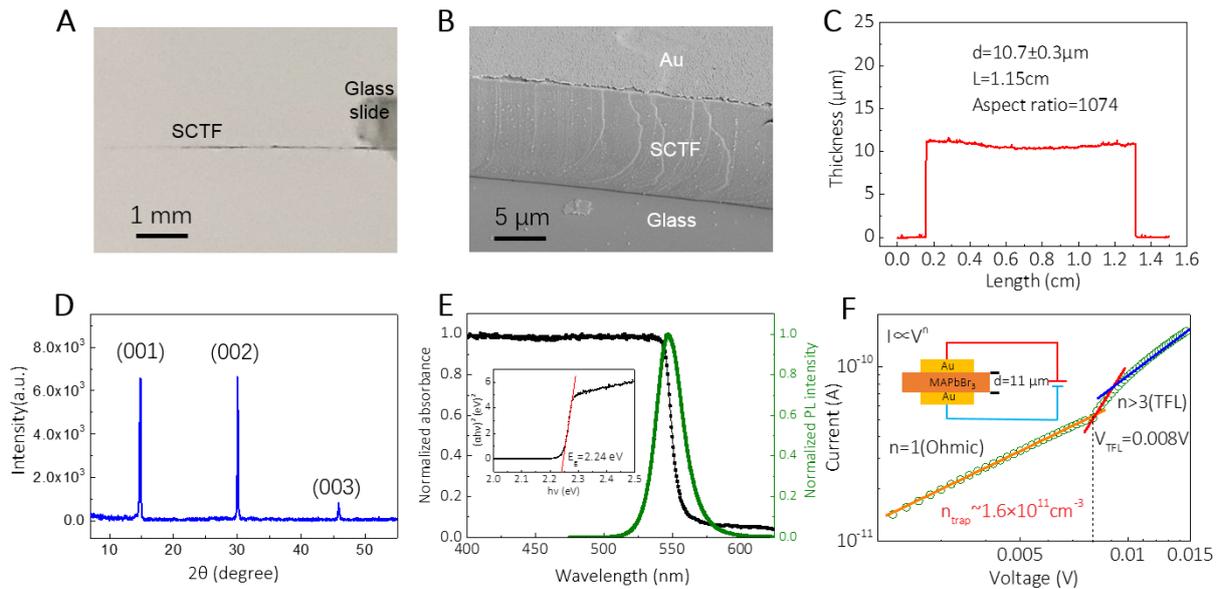

**Figure 3.** Characterization of perovskite SC-TFs. (A) Cross-sectional optical photograph of the SC-TF. (B) Cross-sectional SEM photograph of the SC-TF. (C) Step profiler data of the SC-TF. The thickness of the SC-TF is 10.7 ± 0.3 μm and the length is 1.15 cm, giving an aspect ratio of 1074. (D) X-ray diffraction spectrum of the SC-TF. (E) Absorption spectrum and steady-state PL spectrum of the prepared MAPbBr$_3$ SC-TF. Insets: absorptance versus energy plot of the film, determining the optical bandgap as 2.24 eV. (F) The dark I-V curve used for the trap-state-density measurement. (E, F) Reproduced with permission from Ref. [7], © Springer Nature 2020.

**Conclusions**

In this work, by introducing hydrophobic growth substrates to reduce the adhesion between SCTFs and growth substrates, centimeter-sized perovskite SCTFs can be peeled off from the growth substrates and transferred onto arbitrary target substrates via the anti-solvent intercalation technique. Due to the insolubilization of anti-solvent to perovskite, and the intercalation effect will not form significant stress. The large-scale perovskite SCTFs avoid the chemical corrosion and physical fragmentation in this method, thus maintaining high crystalline quality and original morphology after the transfer process. This approach paves a new way of optimizing perovskite transfer and widens the applications of perovskite SCTFs in large-scale hybrid multilayer optoelectronic devices.

**Acknowledgements:** None.

**Conflict of interest:** The authors declare no competing financial interest.